\documentclass[12pt]{article}
\usepackage{amssymb}
\newcommand{\N}{\mathbb{N}}
\newcommand{\R}{\mathbb{R}}

\newcommand{\Z}{\mathbb{Z}}
\newcommand{\vp}{\tilde{\varphi}}
\newtheorem{theorem}{Theorem}
\newtheorem{proposition}{Proposition}
\newtheorem{lemma}{Lemma}
\title{Massive Scalar Field in an One-Dimensional Oscillating
Region} \author{J.~Dittrich\thanks{Also member of the Doppler
Institute of Mathematical Physics, Faculty of Nuclear Sciences and
Physical Engineering, Czech Technical University, Prague.}\\
{\small\em Nuclear Physics Institute, Academy of Sciences of the
Czech Republic}\\ {\small\em CZ-250 68  \v{R}e\v{z}, Czech
Republic; e-mail dittrich@ujf.cas.cz}\\ \\ P.~Duclos\\ {\small\em
Centre de Physique Th\'{e}orique\thanks{Unite Propre de Recherche
7061.}, CNRS Luminy}\\ {\small\em Case 907, F-13288 Marseille -
Cedex 9, France}\\ {\small\em and}\\ {\small\em Phymat,
Universit\'{e} de Toulon et du Var}\\ {\small\em La Garde, France;
e-mail duclos@univ-tln.fr} } \begin{document} \maketitle

{\abstract{The classical scalar massive field satisfying the Klein-Gordon
equation in a finite one-dimensional space interval of periodically
varying length with Dirichlet boundary conditions is studied. For the
sufficiently small mass, the energy
can exponentially grow with time under the same conditions as for
the massless case. The proofs are based on estimates of exactly
given mass-induced corrections to the massless case.}}

\section{Introduction}
The classical massless scalar field satisfying d'Alembert equation in 1+1
dimensional space-time restricted to finite space interval with
one end-point fixed and other end-point periodically oscillating
was studied in several papers
\cite{Cooper} - \cite{PetrovLlave}. At the end-points, Dirichlet
boundary conditions are assumed. Results for Neumann boundary
conditions are also known \cite{DDG} but the physical
condition on the moving boundary obtained from the static Neumann
one by Lorentz transformation is different and the results for
this condition are more similar to those for Dirichlet boundary
condition. Various regimes of the energy time evolution are
possible - the energy can be unlimited or bounded, with limit or
without limit at time infinity. In particular, there are cases
where energy $E(t)$ is exponentially growing with time $t$ in the
sense that
$$
Ae^{\gamma t} \leq E(t) \leq Be^{\gamma t}
$$
for all $t > 0$ with some $A, B, \gamma > 0$  (the energy is not
monotone in time for a periodic wall motion of course).

Similar model in quantum field theory was also studied in
\cite{Dodonov} and references therein.

In the present paper we consider classical massive scalar field
satisfying Klein-Gordon equation with Dirichlet boundary
conditions at the end-points of finite one-dimensional space
interval, one end-point periodically moving. We prove that the
simplest sufficient conditions for the exponential growth of
the energy are the same as for the d'Alembert equation provided
that the mass is small enough. This shows that the mass can be
considered as a small perturbation.

\section{The Model}
Let $\varphi$ be a real function defined on the space-time region
\begin{equation}
\Omega=\{(t,x)\in\R^2|0\leq x \leq a(t), t\geq 0\}
\end{equation}
where $a:\R\to\R$ is a strictly positive periodic C$^2$-function with a period
$T > 0$. We assume that $|\dot{a}(t)| < 1$ (subluminal velocity
of the end-point motion). We restrict ourselves to the fields
$\varphi$ which are in
C$^2(\buildrel{\,\circ}\over \Omega)\cap$C$^1(\Omega)$ and assume that
Klein-Gordon equation
\begin{equation}
\frac{\partial^2\varphi}{\partial t^2} -
\frac{\partial^2\varphi}{\partial x^2} +
m^2\varphi = 0
\end{equation}
is satisfied in the interior $\buildrel{\,\circ}\over \Omega$. The
constant mass $m$ is non-negative. On the boundary, the conditions
\begin{equation}
\label{cd}
\varphi(0,x)=\varphi_0(x), \quad
\frac{\partial\varphi(0,x)}{\partial t}=\varphi_1(x)
\quad (0<x<a(0))
\end{equation}
\begin{equation}
\label{bc}
\varphi(t,0)=\varphi(t,a(t))=0
\quad (t\geq 0)
\end{equation}
are required.

Let us define
\begin{equation}
\label{hkFdef}
h={\rm Id} - a, \quad k={\rm Id} + a, \quad F=k\circ h^{-1}
\end{equation}
as in \cite{DDG}. Then $h,k,F: \R\to\R$ are increasing
C$^2$-functions.
The identities
\begin{eqnarray}
\nonumber
F={\rm Id}+2a\circ h^{-1}=2h^{-1}-{\rm Id} \quad , \\
\label{Faid}
F^{-1}={\rm Id}-2a\circ k^{-1}=2k^{-1}-{\rm Id} \quad .
\end{eqnarray}
are useful in some calculations.

We shall rewrite the problem in new variables
\begin{equation}
\xi=t+x, \quad \eta=t-x \quad .
\end{equation}
It is not difficult to see that $\Omega$ is transformed into the
set $\tilde{\Omega}$ described by inequalities
\begin{equation}
\label{bc1}
|\eta| \leq \xi \leq F(\eta), \quad \eta \geq -a(0)
\end{equation}
or equivalently
\begin{equation}
\label{bc2}
\max\{-\xi, F^{-1}(\xi)\} \leq \eta \leq \xi, \quad
\xi \geq 0
\end{equation}
the last inequalities in (\ref{bc1}) and (\ref{bc2}) following
automatically from the first ones in fact.
The inequalities describing $\tilde{\Omega}$
can be also written distinguishing two cases
\begin{eqnarray}
0\leq \xi \leq a(0),\quad -\xi\leq\eta\leq\xi \quad ,
\\
\xi\geq a(0), \quad F^{-1}(\xi) \leq \eta \leq \xi \quad .
\end{eqnarray}

The transformed field
\begin{equation}
\vp(\xi,\eta)=\varphi(t,x)
\end{equation}
satisfies the equation
\begin{equation}
\label{KGxi}
\frac{\partial^2\vp}{\partial\xi \partial\eta}=-\frac{m^2}{4}\vp
\end{equation}
in $\buildrel{\,\circ}\over{\tilde{\Omega}}$ with boundary conditions given by
the transformation of equations (\ref{cd}-\ref{bc}).

The energy of the field
\begin{eqnarray}
\label{energydef}
E_m(t)&=&\frac{1}{2}\int_0^{a(t)}\left[
\left(\frac{\partial\varphi(t,x)}{\partial t}\right)^2 +
\left(\frac{\partial\varphi(t,x)}{\partial x}\right)^2 +
m^2\varphi(t,x)^2\right]\,dx \\
\nonumber
&=&\int_0^{a(t)}\left[
\left(\frac{\partial\vp(\xi,\eta)}{\partial \xi}\right)^2 +
\left(\frac{\partial\vp(\xi,\eta)}{\partial \eta}\right)^2 +
\frac{1}{2}m^2\vp(\xi,\eta)^2\right]\,dx \quad .
\end{eqnarray}

\section{Solution of Inhomogeneous Equation}
As a preliminary step, let us consider the equation
\begin{equation}
\label{nhe}
\frac{\partial^2\vp}{\partial\xi \partial\eta}=f(\xi,\eta)
\end{equation}
with given function $f\in$C$^1(\tilde{\Omega})$ for an unknown function $\vp$
satisfying the same boundary conditions as the required
solution of Klein-Gordon equation. We shall put
$f=-\frac{1}{4}m^2\vp$ after some calculations.
The results of this and the next section do not require the
periodicity of function $a$ so it is not assumed here.

We put Equation (\ref{nhe}) into an integral form integrating
twice. The integration bounds must be taken in such a way that
the integration domain is included in $\tilde{\Omega}$. A
possible choice is
\begin{equation}
\frac{\partial\vp(\xi,\eta)}{\partial\xi}=-\int^\xi_\eta f(\xi,z)
dz + H_1(\xi)
\end{equation}
with an arbitrary function $H_1$ (which must be in C$^1$ if $\vp$
should be in C$^2$ of course). It is clear from (\ref{bc2}) that
if $(\xi,\eta)\in\tilde{\Omega}$ then $(\xi,z)\in\tilde{\Omega}$
for all $\eta\leq z\leq\xi$ so the choice of integration bounds
is possible.

Similarly,
\begin{equation}
\vp(\xi,\eta)=
\int^\xi_{|\eta|}\frac{\partial\vp(y,\eta)}{\partial y} dy +
G_1(\eta)
\end{equation}
taking into account (\ref{bc1}). Here $G_1$ is again an arbitrary
function with continuous second derivative (possibly with
exception of the point $\eta=0$).
The last two formulas give
\begin{equation}
\label{nhi}
\vp(\xi,\eta)=-\int^\xi_{|\eta|}dy\int^y_\eta dz f(y,z) + H(\xi) +
G(\eta)
\end{equation}
where $H$ and $G$ are up to now arbitrary functions (such that
$\vp$ is in C$^2$) which has to be determined form the boundary
conditions and Cauchy data.

For $t=0$, i.e. $\xi=-\eta=x\in [0,a(0)]$, Eqs. (\ref{cd}) read
\begin{eqnarray}
\label{cd1}
\varphi_0(\xi)=H(\xi)+G(-\xi)
\\
\label{cd2}
\varphi_1(\xi)=-2\int^\xi_{-\xi} f(\xi,z)dz +H'(\xi)+G'(-\xi)
\end{eqnarray}
Dirichlet boundary condition at $x=0$, i.e. $\xi=\eta=t\geq 0$,
gives
\begin{equation}
\label{dc1}
H(\xi)=-G(\xi)
\end{equation}
and that at $x=a(t)$, i.e. $\xi=F(\eta)$, $\eta\geq -a(0)$, reads
\begin{equation}
\label{dc2}
0=-\int^{F(\eta)}_{|\eta|}dy \int^y_\eta dz f(y,z) + H(F(\eta)) +
G(\eta) \quad .
\end{equation}
By (\ref{dc1}) we can use function $G$ only,
\begin{equation}
\label{phi}
\vp(\xi,\eta)=-\int^\xi_{|\eta|}dy\int^y_\eta dz f(y,z) +
G(\eta) - G(\xi) \quad .
\end{equation}
Equations (\ref{cd1}-\ref{cd2}) give
\begin{equation}
\label{GI}
G(\eta)=-\frac{1}{2}\varphi_0(|\eta|) {\rm sgn}(\eta) -
\frac{1}{2}\int_0^{|\eta|}\varphi_1(x) dx -
\int^{|\eta|}_0 dy \int_{-y}^y dz f(y,z) + c
\end{equation}
for $\eta\in[-a(0),a(0)]$. Here $c$ is an arbitrary constant
which we can choose as zero since $\vp$ is independent of it.
Relation (\ref{dc2}) gives a prolongation formula for $G$,
\begin{equation}
\label{prolong}
G(F(\eta))=G(\eta) - \int^{F(\eta)}_{|\eta|}dy \int^y_\eta dz f(y,z)
\end{equation}
for $\eta\geq -a(0)$. By Eqs. (\ref{phi}-\ref{prolong}), $\vp$ is
determined in the whole domain $\tilde{\Omega}$.

It remains to prove that the above relations really determine a
C$^2$-function $\vp$. The only points where continuity of $\vp$
and its derivatives requires a special check are $\eta=0$,
$\eta=a(0)$ (or $\xi=a(0)$) and their images by the function $F$
(where the continuity then follows automatically). This can be
done by a straightforward but a little tedious calculations which
reveal a sufficient and necessary conditions on the Cauchy data
$\varphi_0$, $\varphi_1$ and the right-hand side function $f$.
The necessity of these conditions can be also easily seen from the
boundary conditions and their derivatives. It is seen that $G$
and $G'$ are continuous in $[-a(0),\infty)$ while $G''$ is
discontinuous at $\eta=0$ and continuous in $[-a(0),0)\cup
(0,\infty)$. We summarize the obtained consistency conditions in
the following Proposition. The derivatives in closed sets are
considered as derivatives with respect to these sets here.
\begin{proposition}
Let $a\in\rm{C}^2([0,\infty))$, $\inf a>0$, $|a'|<1$,
$f\in\rm{C}^1(\tilde{\Omega})$, $\varphi_0\in\rm{C}^2([0,a(0)])$,
$\varphi_1\in\rm{C}^1([0,a(0)])$ and the following relations are
satisfied:
\begin{eqnarray}
\varphi_0(0)=0,\quad \varphi_0(a(0))=0,\\
\varphi_1(0)=0,\quad \varphi_1(a(0))+a'(0)\varphi_0'(a(0))=0,\\
\varphi_0''(0)=-4f(0,0),\\
\nonumber
(1+a'(0)^2)\varphi_0''(a(0))+a''(0)\varphi_0'(a(0))+
2a'(0)\varphi_1'(a(0))\\=-4f(a(0),-a(0)).
\end{eqnarray}
Then there exists a unique $\vp\in\rm{C}^2(\tilde{\Omega})$
satisfying the equation (\ref{nhe}) in $\tilde{\Omega}$ and
boundary conditions corresponding to the transformed relations
(\ref{cd}-\ref{bc}). The solution $\vp$ is given by Eqs.
(\ref{phi}-\ref{prolong}).
\end{proposition}
\hspace{13cm} $\fbox{}$

\section{Existence and Unicity of the Solution}
\begin{figure}
\begin{center}
\setlength{\unitlength}{0.03mm}
\hskip-5cm
\begin{picture}(3500,3500)
\put(1500,500){\line(1,0){500}}
\thicklines\put(2000,500){\line(1,0){1000}}\thinlines
\put(3000,500){\vector(1,0){700}}
\put(3440,560){\small\mbox{$x$}}
\put(2000,0){\line(0,1){500}}
\thicklines\put(2000,500){\vector(0,1){2580}}\thinlines
\put(1900,3000){\small\mbox{$t$}}
\put(1500,0){\line(1,1){840}}
\put(2520,1020){\vector(1,1){980}}
\put(3360,2000){\small\mbox{$\xi$}}
\put(2500,0){\vector(-1,1){1400}}
\put(1000,1300){\small\mbox{$\eta$}}
\thicklines
\bezier{0}(3000,500)(2800, 750)(3000,1000)
\bezier{0}(3000,1000)(3200, 1250)(3000,1500)
\bezier{0}(3000,1500)(2800,1750)(3000,2000)
\bezier{0}(3000,2000)(3200, 2250)(3000,2500)
\bezier{0}(3000,2500)(2800, 2750)(3000,3000)
\thinlines
\put(2620,2940){\line(1,-1){299}}
\put(2919,2641){\line(-1,-1){919}}
\put(2000,1722){\line(1,-1){907}}
\put(2907,815){\line(-1,-1){315}}
\put(2620,2940){\line(-1,-1){620}}
\put(2000,2320){\line(1,-1){1099}}
\put(3099,1221){\line(-1,-1){721}}
\put(2480,3000){\small\mbox{$(\xi,\eta)$}}
\put(2516,2021){\vector(-1,0){190}}
\put(2510,1981){\small\mbox{$B(\xi,\eta)$}}
\put(1685,2285){\small\mbox{$(\eta,\eta)$}}
\put(2950,2610){\small\mbox{$(\xi,F^{-1}(\xi))$}}
\put(2575,922){\vector(1,0){190}}
\put(2115,900){\small\mbox{$B^2(\xi,\eta)$}}
\put(2420,2450){\small\mbox{$-$}}
\put(2540,1430){\small\mbox{$+$}}
\put(2620,650){\small\mbox{$-$}}
\end{picture}
\caption{\em
The set $M(\xi,\eta)$ bounded by backward characteristics
and the signs of function $\vartheta(\xi,\eta,y,z)$. The
coordinates of the points are indicated in variables $\xi$ and
$\eta$. The number $N(\xi,\eta)=2$ for the displayed case.
}
\label{FigMset}
\end{center}
\end{figure}
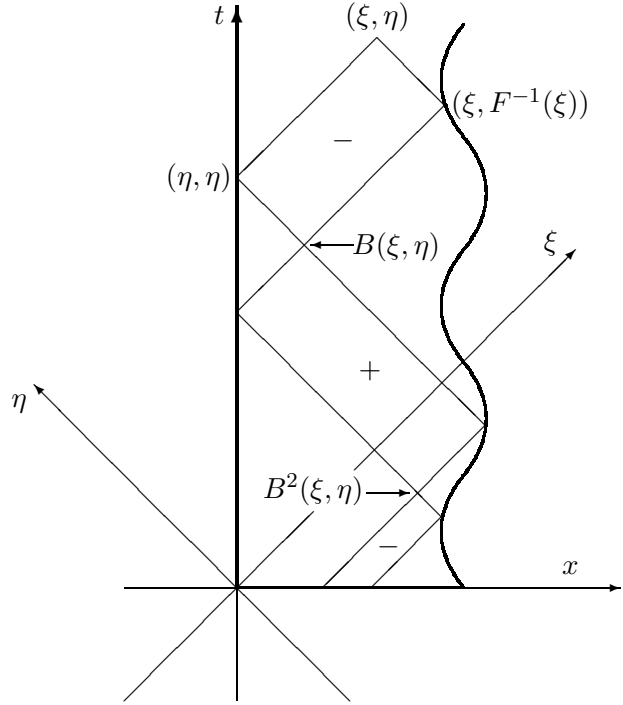

We use the results of previous section to write down the integral
form of the Klein-Gordon equation in our case. Let us start with
some notations. For $(\xi,\eta)\in\tilde{\Omega}$ let us denote
the corresponding time $t$ as
\begin{equation}
\label{Tdef}
T(\xi,\eta)=\frac{\xi+\eta}{2} \quad ,
\end{equation}
\begin{eqnarray}
\nonumber
Q(\xi,\eta)=\{(y,z)\in\R^2| \eta\leq y \leq \xi, F^{-1}(\xi)\leq z \leq
\eta\}\\ = [\eta,\xi]\times [F^{-1}(\xi),\eta]
\label{Qdef}
\end{eqnarray}
the rectangle bounded by backward characteristics starting from
$(\xi,\eta)$ (see Fig.~\ref{FigMset}) and
\begin{equation}
\label{Bdef}
B(\xi,\eta)=(\eta,F^{-1}(\xi))
\end{equation}
its lowest vertex. It is easy to verify that the written formulas
correspond to Fig.~\ref{FigMset}.
The following trivial facts are easily seen.
\begin{lemma}
\label{Qpos}
Let $(\xi,\eta)\in\tilde{\Omega}$. Then
\begin{eqnarray}
(y,z)\in Q(\xi,\eta) \Longrightarrow T(B(\xi,\eta))\leq T(y,z)
\leq T(\xi,\eta) \quad ,\\
\label{QinO}
Q(\xi,\eta)\subset \tilde{\Omega} \Longleftrightarrow
B(\xi,\eta)\in\tilde{\Omega} \Longleftrightarrow T(B(\xi,\eta))\geq
0 \quad ,\\
\nonumber
Q(\xi,\eta)\cap\tilde{\Omega} = \{(y,z)\in Q(\xi,\eta)|
T(y,z)\geq 0\}\\
\label{QO}
=\{(y,z)\in\R^2|\, |\eta|\leq y\leq\xi, \max(F^{-1}(\xi),-y)\leq z
\leq \eta\} \quad ,\\
\label{Tdif}
T(\xi,\eta)-T(B(\xi,\eta))=a\circ k^{-1}(\xi)\geq \inf a \quad .
\end{eqnarray}
\end{lemma}
\hspace{13cm} $\fbox{}$
\begin{lemma}
\label{ieQ}
\begin{description}
\item{(a)}
Let $\vp\in C^2(\tilde{\Omega})$ satisfies Eq. (\ref{KGxi}) and
the Dirichlet boundary conditions corresponding to (\ref{bc}). Then
\begin{equation}
\label{Binstep}
\vp(\xi,\eta)=-\vp(B(\xi,\eta)) -
\frac{m^2}{4}\int_{Q(\xi,\eta)}\vp(y,z)\, dy\, dz \quad
\end{equation}
for every $(\xi,\eta)\in\tilde{\Omega}$ such that $T(\xi,\eta)\geq 0$.
\item{(b)}
Let $(\xi_0,\eta_0)\in\tilde{\Omega}$ be such that $T(B(\xi_0,\eta_0))>0$
and $\vp\in C^0(\tilde{\Omega})$ has continuous first derivatives in a
neighborhood of boundary $\partial Q(\xi_0,\eta_0)$.
If $\vp$ is of
class $C^2$ in a neighborhood of point $B(\xi_0,\eta_0)$,
satisfies Eq. (\ref{KGxi}) there,
and satisfies Eq. (\ref{Binstep})
for $(\xi, \eta)$ in a neighborhood of $(\xi_0,\eta_0)$
then $\vp$ is of class $C^2$ in a neighborhood of
point $(\xi_0,\eta_0)$ and satisfies Eq. (\ref{KGxi}) there.
\end{description}
\end{lemma}
{\bf Proof.}
Notice that $Q(\xi,\eta)\subset\tilde{\Omega}$ for
$T(B(\xi,\eta))\geq 0$ according to (\ref{QinO}).
For the statement (a), integrate Eq. (\ref{KGxi})
over (\ref{Qdef}) and use
the Dirichlet boundary conditions at the vertices $(\xi_1=\eta,\eta)$
and $(\xi,\eta_1=F^{-1}(\xi))$ (notation as in Fig.~\ref{FigMset}). To see
the statement (b), use Eqs. (\ref{Qdef}), (\ref{Bdef}) and differentiate
(\ref{Binstep}). In particular,
$$
\frac{\partial^2\vp (\xi,\eta)}{\partial\xi \partial\eta} =
-\left[\frac{\partial^2\vp}{\partial\xi \partial\eta} +
\frac{m^2}{4}\vp\right]_{\vert B(\xi,\eta)} \frac{d}{d\xi}F^{-1}(\xi)
- \frac{m^2}{4}\vp(\xi,\eta) \, .
$$
\\
\hspace*{13cm} $\fbox{}$\\
{\bf Remark.} For $m=0$ the Lemma is a special case of the well
known relation of four values at the vertices of rectangle bounded
by characteristics for the solution of the wave equation
(e.g. Eq. (1.22) of Chapter 8 in \cite{DAEQref}).
\begin{lemma}
\label{ieQ0}
Let $(\xi,\eta)\in\tilde{\Omega}$ be such that
$T(B(\xi,\eta))\leq 0$, $\vp$ satisfies Eq.
(\ref{KGxi}) with boundary conditions corresponding to
(\ref{cd}), (\ref{bc}) and $\vp^{(0)}$ satisfies Eq. (\ref{KGxi})
with $m$ replaced by zero and the same boundary conditions
(\ref{cd}), (\ref{bc}) as $\vp$. Then
\begin{equation}
\label{phiphi01}
\vp(\xi,\eta)=\vp^{(0)}(\xi,\eta) -
\frac{m^2}{4}\int_{Q(\xi,\eta)\cap\tilde{\Omega}}\vp(y,z)
\, dy\, dz \quad .
\end{equation}
If $\vp^{(0)}\in C^2(\tilde{\Omega})$ and $\vp\in C^0(\tilde{\Omega})$
satisfies the relation (\ref{phiphi01}) then $\vp$ satisfies the
boundary conditions corresponding to (\ref{cd}), (\ref{bc}) and
Eq. (\ref{KGxi}). Further, $\vp$ has continuous derivatives up to
the second order in the considered part of $\tilde{\Omega}$.
\end{lemma}
{\bf Proof.}
Let us first realize that for $(\xi,\eta)$ on the boundary of
$\tilde{\Omega}$ the Lebesgue measure of the set
$Q(\xi,\eta)\cap\tilde{\Omega}$ is zero so the first initial
condition (\ref{cd}) and boundary conditions (\ref{bc})
are satisfied if Eq. (\ref{phiphi01}) holds.

We shall use equations (\ref{phi}-\ref{prolong}) for
the several ranges of variables $\xi$, $\eta$.
We denote as $G_0$ the function (\ref{GI}-\ref{prolong})
corresponding to the d'Alembert equation solution $\vp^{(0)}$
with the zero mass.
As for $T(B(\xi,\eta))\leq 0$
\begin{equation}
\label{TBN}
F^{-1}(\xi)\leq -\eta\leq a(0)
\end{equation}
(see (\ref{Tdef}) and
(\ref{bc1})), $\xi\leq F(a(0))$.
From (\ref{bc2}) and (\ref{TBN}) we see that
$F^{-1}(\xi)\leq\eta\leq -F^{-1}(\xi)$, i.e.
\begin{equation}
\label{etafxi}
F^{-1}(\xi)\leq 0\quad , \quad |\eta|\leq |F^{-1}(\xi)| \quad .
\end{equation}
Further it is clear that $\xi$ and $\eta$ cannot be
both simultaneously greater than $a(0)$ as $T(B(\xi,\eta))$ would
be positive in this case (remember that $F^{-1}(a(0))=-a(0)$).
Taking into account (\ref{bc2}) we see that always $\eta\leq a(0)$
under the assumptions of Lemma.
We have to distinguish two cases now.
\\
\\
\noindent
{\em 1) $\xi\leq a(0)$, $\eta\leq a(0)$}\\
Now equation (\ref{GI}) can be used in (\ref{phi}) and we obtain
\begin{eqnarray}
\nonumber
\lefteqn{\vp(\xi,\eta)=} \\
\nonumber
& & \vp^{(0)}(\xi,\eta) + \frac{m^2}{4} \left\lbrack
\int_{|\eta|}^{\xi}dy\int_{\eta}^{y}dz +
\int_{0}^{|\eta|}dy\int_{-y}^{y}dz - \int_{0}^{\xi}dy\int_{-y}^{y}dz
\right\rbrack \vp(y,z) \\
\nonumber
& & =  \vp^{(0)}(\xi,\eta) + \frac{m^2}{4} \left\lbrack
\int_{|\eta|}^{\xi}dy\int_{\eta}^{y}dz -
\int_{|\eta|}^{\xi}dy\int_{-y}^{y}dz \right\rbrack \vp(y,z) \\
\nonumber
& & =  \vp^{(0)}(\xi,\eta) - \frac{m^2}{4}
\int_{|\eta|}^{\xi}dy\int_{-y}^{\eta}dz\, \vp(y,z)
\end{eqnarray}
which is (\ref{phiphi01}) due to (\ref{QO}) as here
$F^{-1}(\xi)\leq-a(0)\leq-\xi\leq-y$. On the contrary,
differentiating the last equation and using the already proved
boundary conditions we can verify that $\vp$ has continuous second
derivatives and satisfies (\ref{KGxi}) if the integral relation
(\ref{phiphi01}) holds. The initial condition for the time
derivative in (\ref{cd}) is also seen.
\\
\\
\noindent
{\em 2) $a(0)<\xi\leq F(a(0))$, $\eta\leq a(0)$}\\
Now formula (\ref{GI}) holds for $G(\eta)$ while (\ref{prolong})
gives
\begin{eqnarray*}
\lefteqn{G(\xi)=G(F^{-1}(\xi)) +
\frac{m^2}{4}\int_{|F^{-1}(\xi)|}^\xi dy\int_{F^{-1}(\xi)}^{y}dz
\,\vp(y,z)}\\ & &
=G_0(\xi) +
\frac{m^2}{4}\left\lbrack
\int_{0}^{|F^{-1}(\xi)|}dy\int_{-y}^{y}dz  +
\int_{|F^{-1}(\xi)|}^\xi dy\int_{F^{-1}(\xi)}^{y}dz\right\rbrack
\vp(y,z)
\end{eqnarray*}
and
\begin{eqnarray*}
\lefteqn{
\vp(\xi,\eta)= \vp^{(0)}(\xi,\eta) + \frac{m^2}{4} \left\lbrack
\int_{|\eta|}^{\xi}dy\int_{\eta}^{y}dz +
\int_{0}^{|\eta|}dy\int_{-y}^{y}dz \right. }\\
& & \left. - \int_{0}^{|F^{-1}(\xi)|}dy\int_{-y}^{y}dz\  -
\int_{|F^{-1}(\xi)|}^\xi dy\int_{F^{-1}(\xi)}^{y}dz
\right\rbrack \vp(y,z) \quad .
\end{eqnarray*}
With the help of (\ref{etafxi}) we combine the second and third
integral and the first and fourth integral separating the first
one into two parts. Then
\begin{eqnarray*}
\lefteqn{\vp(\xi,\eta)= \vp^{(0)}(\xi,\eta) +  \frac{m^2}{4} \left\lbrack
\int_{|\eta|}^{|F^{-1}(\xi)|}dy\int_{\eta}^{y}dz
\right. } \\
& & \left. -
\int_{|F^{-1}(\xi)|}^\xi dy\int_{F^{-1}(\xi)}^{\eta}dz
-\int_{|\eta|}^{|F^{-1}(\xi)|}dy\int_{-y}^{y}dz
\right\rbrack \vp(y,z)\\
&=& \vp^{(0)}(\xi,\eta) +  \frac{m^2}{4} \left\lbrack
-\int_{|\eta|}^{|F^{-1}(\xi)|}dy\int_{-y}^{\eta}dz -
\int_{|F^{-1}(\xi)|}^\xi dy\int_{F^{-1}(\xi)}^{\eta}dz
\right\rbrack \vp(y,z)\\
&=&
 \vp^{(0)}(\xi,\eta) -  \frac{m^2}{4}
\int_{|\eta|}^{\xi}dy\int_{\max(F^{-1}(\xi),-y)}^{\eta}dz
\,\vp(y,z)\\
\end{eqnarray*}
which is (\ref{phiphi01}).
On the contrary,
differentiating the next-to-last equation and using the already proved
boundary conditions we can verify that $\vp$ has continuous second
derivatives and satisfies (\ref{KGxi}) if the integral relation
(\ref{phiphi01}) holds.

The continuity of the first and second derivatives at $\xi=a(0)$
can be also seen.
\nopagebreak
\\
\nopagebreak
\hspace*{13cm} $\fbox{}$\\

Let us denote the iterations of the map $B$ as
$$
B^0(\xi,\eta)=(\xi,\eta),\; B^1(\xi,\eta)=B(\xi,\eta),\;
B^2(\xi,\eta)=B(B(\xi,\eta)),\; \dots
$$
and let
\begin{equation}
\label{Ndef}
N(\xi,\eta)=\max\{n\in\Z|\, B^n(\xi,\eta)\in\tilde{\Omega}\}
\quad ,
\end{equation}
\begin{equation}
\label{Mdef}
M(\xi,\eta)=\bigcup_{n=0}^{N(\xi,\eta)-1}Q(B^n(\xi,\eta))\,\cup\,
(Q(B^{N(\xi,\eta)}(\xi,\eta))\cap\tilde{\Omega})
\end{equation}
(see Fig.~\ref{FigMset}),
\begin{equation}
\vartheta(\xi,\eta,y,z)=(-1)^{n+1}
\end{equation}
for $(y,z)\in Q(B^n(\xi,\eta))$, $n\in\Z$.
\begin{theorem}
\label{ie}
Let
$\vp^{(0)}$ satisfies Eq. (\ref{KGxi})
with $m$ replaced by zero and boundary conditions corresponding
to (\ref{cd}), (\ref{bc}). If
$\vp\in C^2(\tilde{\Omega})$ satisfies Eq.
(\ref{KGxi}) with boundary conditions corresponding to
(\ref{cd}), (\ref{bc}) then
\begin{equation}
\label{phiphi0}
\vp(\xi,\eta)=\vp^{(0)}(\xi,\eta) +
\frac{m^2}{4}\int_{M(\xi,\eta)}\vartheta(\xi,\eta,y,z)\vp(y,z)
\, dy\, dz \quad .
\end{equation}
Conversely, if $\vp\in C^0(\tilde{\Omega})$ satisfies Eq.
(\ref{phiphi0}) then $\vp\in C^2(\tilde{\Omega})$, $\vp$
satisfies Eq. (\ref{KGxi}) and the boundary conditions
corresponding to (\ref{cd}-\ref{bc}).
\end{theorem}
{\bf Proof.} Let $(\xi,\eta)\in\tilde{\Omega}$. By Eq.
(\ref{Tdif}), $N(\xi,\eta)$ defined by (\ref{Ndef}) exists
since $\inf\{a(t)\;|\;0\leq t\leq T(\xi,\eta)\}>0$.
With the help
of Lemmas \ref{ieQ} and \ref{ieQ0}, the Theorem follows by
induction with respect to $N(\xi,\eta)$. The continuity of $\vp$
and its derivatives up to the second order
at the curve $T(B(\xi,\eta))=0$ separating the regions of
validity of the two lemmas can be also checked.
\\
\hspace*{13cm} $\fbox{}$\\
\begin{lemma}
\label{Mmeasure}
Let $a_{max}:=\sup a < \infty$ and $(\xi,\eta)\in\tilde{\Omega}$.
Then
\begin{equation}
\label{Mestim}
M(\xi,\eta)\subset\{(y,z)\in\R^2|0\leq y\leq \xi,\,
y-2 a_{max} \leq z \leq y \}
\end{equation}
and the Lebesgue measure of the set $M(\xi,\eta)$ verifies
\begin{equation}
\label{Mmestim}
0\leq \int_{M(\xi,\eta)}\,dy\,dz\,\leq 2 a_{max} T(\xi,\eta)
\leq\ 2 a_{max} \xi \quad .
\end{equation}
If $(\xi,\eta)$ is on the boundary of $\tilde{\Omega}$ then the
Lebesgue measure of $M(\xi,\eta)$ is zero.
\end{lemma}
{\bf Proof.}
Let $(y,z)\in M(\xi,\eta)$. By Eq. (\ref{Qdef}), $y\leq\xi$ for
$(y,z)\in Q(\xi,\eta)$. Looking at Eqs. (\ref{Mdef}),
(\ref{Bdef}) and (\ref{bc2}) the inequality $y\leq\xi$ is seen
for all $(y,z)\in M(\xi,\eta)$. Further $y\geq 0$ as
$M(\xi,\eta)\subset \tilde{\Omega}$ (see again (\ref{bc2})). By
(\ref{Qdef}), $z\leq \eta \leq y$ and
$$
z\geq F^{-1}(\xi)\geq F^{-1}(y) = y - 2a\circ k^{-1}(y) \geq
y - 2 a_{max}
$$
for arbitrary $(y,z)\in Q(\xi,\eta)$,
$(\xi,\eta)\in\tilde{\Omega}$.
The estimate (\ref{Mmestim}) now immediately follows using
variables $t$ and $x$ to calculate the first bound.

The boundary of $\tilde{\Omega}$ consists of points for which
\begin{description}
\item{(i)}
$\xi=-\eta$, i.e. $t=0$,
\item{(ii)}
$\xi=\eta$, i.e. $x=0$,
\item{(iii)}
$\xi=F(\eta)$, i.e. $x=a(t)$.
\end{description}
In the case (i), $M(\xi,\eta)$ is one-point and
therefore zero-measure. In the cases (ii) and (iii), the measure
of $Q(\xi,\eta)$ is zero by (\ref{Qdef}) and $B(\xi,\eta)$ is
also on the boundary of $\tilde{\Omega}$ if still
$B(\xi,\eta)\in\tilde{\Omega}$. So the measure of
$M(\xi,\eta)$ is zero according to (\ref{Mdef}).
\\
\hspace*{13cm} $\fbox{}$\\
\begin{theorem}
\label{KGEexist}
Let $a\in\rm{C}^2([0,\infty))$, $0<\inf a\leq\sup a<\infty$, $|a'|<1$,
$\varphi_0\in\rm{C}^2([0,a(0)])$,
$\varphi_1\in\rm{C}^1([0,a(0)])$ and the following relations are
satisfied:
\begin{eqnarray}
\varphi_0(0)=0,\quad \varphi_0(a(0))=0,\\
\varphi_1(0)=0,\quad \varphi_1(a(0))+a'(0)\varphi_0'(a(0))=0,\\
\varphi_0''(0)=0,\\
(1+a'(0)^2)\varphi_0''(a(0))+a''(0)\varphi_0'(a(0))+
2a'(0)\varphi_1'(a(0))=0.
\end{eqnarray}
Then there exists unique $\vp\in\rm{C}^2(\tilde{\Omega})$
satisfying the equation (\ref{KGxi}) in $\tilde{\Omega}$ and
boundary conditions corresponding to the transformed relations
(\ref{cd}-\ref{bc}). The solution satisfies the estimate
\begin{equation}
\label{phibound}
|\vp(\xi,\eta)|\leq c e^{\frac{1}{2}a_{max} m^2 \xi}
\end{equation}
with a constant $0<c<\infty$ independent of $m$ (but dependent on
$\varphi_0$ and $\varphi_1$).
\end{theorem}
{\bf Proof.}
We iterate the Equation (\ref{phiphi0}) denoting
\begin{eqnarray}
\label{iter1}
\vp^{(n)}(\xi,\eta)&=&\vp^{(0)}(\xi,\eta) +
\frac{m^2}{4}\int_{M(\xi,\eta)}\vartheta(\xi,\eta,y,z)\vp^{(n-1)}(y,z)
\, dy\, dz \; ,
\\
\label{iter2}
\varepsilon^{(n-1)}(\xi,\eta)&=&\vp^{(n)}(\xi,\eta) -
\vp^{(n-1)}(\xi,\eta)
\end{eqnarray}
for $n=1,2,3,\dots$. Now
\begin{equation}
\label{iter3}
\varepsilon^{(n)}(\xi,\eta)=
\frac{m^2}{4}\int_{M(\xi,\eta)}\vartheta(\xi,\eta,y,z)
\varepsilon^{(n-1)}(y,z) \, dy\, dz \quad .
\end{equation}
Under our assumptions, all the functions $\vp^{(n)}$ and
$\varepsilon^{(n)}$ are continuous. Then the estimate
\begin{equation}
\label{epsest}
\left|\varepsilon^{(n)}(\xi,\eta)\right| \leq
\frac{a_{max}^n m^{2n} \xi^n}{2^n\, n!}
\left\|\varepsilon^{(0)}\right\|_{L^{\infty}(M_0)}
\end{equation}
follows by induction using Lemma \ref{Qpos} and Lemma \ref{Mmeasure}
for
$$
(\xi,\eta)\in M_0:=\{(y,z)\in\tilde{\Omega}\,|\, T(y,z)\leq T_0\}
$$
with arbitrary given $T_0>0$.
So the sequence
\begin{equation}
\vp^{(n+1)}=\sum_{k=0}^n\varepsilon^{(k)} + \vp^{(0)}
\end{equation}
is uniformly convergent in any compact subset of
$\tilde{\Omega}$, its limit $\vp$ is continuous, satisfies
Equation (\ref{phiphi0}) and the required boundary and initial
conditions. Therefore $\vp$ satisfies also Equation
(\ref{KGxi}).

Assume that we have two solutions of Equation (\ref{KGxi})
satisfying the required boundary and initial conditions. Then
they satisfy also Equation (\ref{phiphi0}) and the estimate like
(\ref{epsest}) holds for their difference with any $n$. So the
two solutions must be identical and the uniqueness is proved.

The solution $\vp^{(0)}$ of d'Alembert equation is known to be
bounded in $\tilde{\Omega}$ (see also (\ref{phi}) and
(\ref{prolong}) for $f=0$). Equations
(\ref{iter1}) then leads to
\begin{equation}
\label{phibound0}
|\vp(\xi,\eta)|\leq \|\vp^{(0)}\|_\infty
e^{\frac{1}{2}a_{max}m^2\xi}
\end{equation}
and (\ref{phibound}) is proved.
\\
\hspace*{13cm} $\fbox{}$

\section{Energy Large-Time Behavior}
In this section we are going to prove that the energy can be
exponentially increasing (up to non-monotone evolution within
the period of the end-point motion) for sufficiently small mass
under the same assumptions as for the massless case. Let us first
write a formula for the function $G$ by iterations of relation
(\ref{prolong}). Let $G_0$ be the corresponding function for
$m=0$. For any $\eta\in [-a(0),\infty)$ there exists just one
non-negative integer $n(\eta)$ such that
\begin{equation}
\label{ndef}
\eta\in F^{n(\eta)}([-a(0),a(0)))
\quad .
\end{equation}
We shall use also integer
\begin{equation}
\label{Ktdef}
K(t)=n(t+a_{max})\geq n(\xi)\geq n(\eta) \quad .
\end{equation}
By induction with respect to $n(\eta)$ and comparison of the
relations with $m=0$ and $m\not= 0$ we obtain
\begin{eqnarray}
\nonumber
G(\eta)=G_0(\eta) &+&
\frac{m^2}{4}\int_0^{|F^{-n(\eta)}(\eta)|}\,dy\,\int_{-y}^y\,dz\,\vp(y,z)
\\ &+&
\frac{m^2}{4}\sum_{j=1}^{n(\eta)}\int_{|F^{-j}(\eta)|}^{F^{1-j}(\eta)}
\,dy\,\int_{F^{-j}(\eta)}^y\,dz\,\vp(y,z)\; .
\end{eqnarray}
To calculate the energy density, we need also derivatives of the
function $\vp$ and therefore $G$. Taking into account that
$F^{-j}(\eta)$ can be negative only for $j=n(\eta)$ and excluding
a discrete set of values of $\eta$ we obtain
\begin{eqnarray}
\nonumber
G'(\eta) &=& G_0'(\eta)
\\ \nonumber
&+& \frac{m^2}{4}\Biggl\{\frac{dF^{-n(\eta)}(\eta)}{d\eta}
{\rm sgn}\left(F^{-n(\eta)}(\eta)\right) \cdot
\\ \nonumber
&\cdot& \int_{-|F^{-n(\eta)}(\eta)|}
^{\delta_{n(\eta),0}|F^{-n(\eta)}(\eta)| +
(1-\delta_{n(\eta),0})F^{-n(\eta)}(\eta)}
\vp(|F^{-n(\eta)}(\eta)|,z)\,dz
\\ \nonumber
&+& \sum_{j=1}^{n(\eta)}\frac{dF^{1-j}(\eta)}{d\eta}
\int_{F^{-j}(\eta)}^{F^{1-j}(\eta)}\vp(F^{1-j}(\eta),z)\,dz
\\
&-& \sum_{j=1}^{n(\eta)}\frac{dF^{-j}(\eta)}{d\eta}
\int_{|F^{-j}(\eta)|}^{F^{1-j}(\eta)}\vp(y,F^{-j}(\eta))\,dy
\Biggr\} \quad .
\end{eqnarray}
This formula has the form
\begin{equation}
G'(\eta) = G_0'(\eta) +
\frac{m^2}{4}\sum_{j=0}^{n(\eta)}B_j(\eta)\frac{dF^{-j}(\eta)}{d\eta}
\end{equation}
where $B_j(\eta)$ are seen above. Using (\ref{phibound}) and the
relation
\begin{equation}
\label{Fstep}
\eta - F^{-1}(\eta) = 2a\circ k^{-1}(\eta) \leq 2a_{max}
\end{equation}
following from (\ref{Faid}), the estimate
\begin{equation}
\label{Bestim}
\left| B_j(\eta) \right| \leq c_1(m)
e^{\frac{1}{2}a_{max}m^2 |F^{-j}(\eta)|}
\end{equation}
is shown for $j=0,\dots,n(\eta)>0$ with
\begin{equation}
c_1(m)=2 a_{max} c \left( 1 + e^{a_{max}^2m^2} \right) \quad .
\end{equation}
We indicated here the dependence of constant $c_1$ on the mass
$m$ but we do not indicate the automatically assumed dependence on
the initial data $\varphi_0$, $\varphi_1$ (see (\ref{phibound}) and
(\ref{phibound0})) and the function $a$. We shall keep such
notation for further constants in estimates below as we finally
want to have an $m$-independent estimate over the range of mass
values.
By (\ref{ndef}) and (\ref{Fstep}),
\begin{equation}
\label{Fmkup}
|F^{-j}(\eta)|\leq [2n(\eta)-2j+1] a_{max}
\end{equation}
for $j=0,\dots,n(\eta)$ and therefore
\begin{equation}
\left| B_j(\eta) \right| \leq c_1(m)
e^{(n(\eta)-j+\frac{1}{2})a_{max}^2m^2} \quad .
\end{equation}
Using the above formulas together with Eq. (\ref{phi}) we obtain
for $\omega=\xi, \eta$
\begin{equation}
\label{deriter}
\frac{\partial\vp(\xi,\eta)}{\partial\omega} =
\frac{\partial\vp^{(0)}(\xi,\eta)}{\partial\omega} +
\frac{m^2}{4} \sum_{j=0}^{n(\omega)}A_j^{(\omega)}(\xi,\eta)
\frac{d F^{-j}(\omega)}{d \omega}
\end{equation}
where
\begin{eqnarray}
A_0^{(\xi)}(\xi,\eta) &=& \int_\eta^\xi\vp(\xi,z)\,dz - B_0(\xi)
\quad , \quad A_j^{(\xi)}(\xi,\eta) = - B_j(\xi) \quad ,
\\
A_0^{(\eta)}(\xi,\eta) &=&
-{\rm sgn}(\eta)\int_\eta^{|\eta|}\vp(|\eta|,z)\,dz -
\int_{|\eta|}^\xi\vp(y,\eta)\,dy + B_0(\eta) \; , \\
A_j^{(\eta)}(\xi,\eta) &=& B_j(\eta)
\end{eqnarray}
for $j=1,\dots,n(\omega)$. The upper estimate of
$2ca_{max}e^{\frac{1}{2}a_{max}m^2\xi}$ for integral terms in both formulas
for $A_0^{(\omega)}(\xi,\eta)$ can be seen from (\ref{phibound}).
Realizing that $n(\eta)\leq n(\xi)$ as $\eta\leq \xi$ and  that
$\xi<(2n(\xi)+1)a_{max}$ according to (\ref{Fmkup}), we arrive at
\begin{eqnarray}
\label{Aestim}
|A_j^{(\omega)}(\xi,\eta)| &\leq& c_2(m) e^{(n(\xi)-j)a_{max}^2m^2} \quad ,
\\
c_2(m) &=&
2 a_{max} c \left( 2 + e^{a_{max}^2m^2} \right)
e^{\frac{1}{2}a_{max}^2m^2}
\end{eqnarray}
for $j=0,\dots,n(\omega)$ and $\omega=\xi,\eta$.

To estimate the contributions of terms in (\ref{deriter}) to the
energy (\ref{energydef})
we shall use the results for the d'Alembert equation
\cite{Cooper, DDG}.
Let us first remind them in a
form suitable for that.
Function $F$ defined in (\ref{hkFdef}) is an increasing
diffeomorphism of the line $\R$ satisfying the relation $F(t+T)=F(t)+T$
for $t\in R$, i.e. a covering of a diffeomorphism of the circle of
length $T$.
The notions of the rotation number  and
periodic point used
below are defined e.g. in \cite{Herman}, a brief review is given
also in \cite{DDG}. We shall use the notation
$$
F^n=F \circ \dots \circ F
$$
for the composition of the function $F$ $n$-times with itself,
$(f)^n$ for the $n$-th power of the function $f$, i.e. for the
function with values
$$
f(x)^n=f(x)\cdot \dots \cdot f(x) \quad ,
$$
$DF=F'$ for the derivative of function $F$, and $\chi_I$ for the
characteristic function of the interval $I$ with the value equal
$1$ in $I$ and $0$ outside $I$.
\begin{lemma}
\label{asymptotpq}
Let function $F$ has the rotation number
$\rho(F)=\frac{p}{q}T$ where $p\in\N^*=\{1,2,\dots\}$ and $q\in\N^*$ are
relatively prime, $a_1$ be an attracting periodic point of $F$,
the function $F$ has in $[a_1,F(a_1))$
a finite number of periodic points of period $q$ from which
$a_1<a_2<\dots<a_N$
are attracting with
$$
DF^q(a_i)<1 \quad {\rm for}\; i=1,\dots,N
$$
and other periodic points in $[a_1,F(a_1))$ are repelling.
Let us denote as $b_{i-1}$, $b_i$ the nearest repelling periodic points
to $a_i$ such that $b_{i-1}<a_i<b_i$, and as
$J_1=[a_1,b_1)\cup(b_N,F(a_1))$, $J_i=(b_{i-1},b_i)$
for $i=2,\dots,N$.
Let $f\in L^2((a_1,F(a_1)))$ be a real function, $\|f\|>0$.
Then
\begin{equation}
\label{nppasympt}
\int_{a_1}^{F(a_1)}\frac{f(x)^2}{DF^{nq}(x)}\,dx =
\sum_{i=1}^N A_i [DF^q(a_i)]^{-n} + R_n
\end{equation}
where $0\leq A_i<\infty$, $A_i>0$ if $\|f\|_{L^2(J_i)}>0$,
\begin{equation}
\label{oneppasympt}
A_i=\|\sqrt{l^{(i)}} f\|^2_{L^2(J_i)}
\quad , \quad
l^{(i)}(x)=\prod_{k=0}^\infty\frac{DF^q(a_i)}{DF^q\circ F^{kq}(x)}
\end{equation}
for $x\in J_i$ and $i=1,\dots,N$. The remainder
\begin{equation}
\label{restasympt}
R_n=o([DF^q(a_{i_0})]^{-n})
\end{equation}
as $n\to\infty$, $i_0$ being defined by the relation
\begin{equation}
DF^q(a_{i_0})=\min\{DF^q(a_i)|\, i=1,\dots,N\;{\rm and}\; A_i>0\}
\quad .
\end{equation}
\end{lemma}
{\bf Proof.}
The set of periodic points $b$ satisfying
$$
F^q(b)=b+pT \quad ,
$$
the set of attracting periodic points, and the set of repelling
periodic points are invariant under the action of function $F$.
Further
$$
DF^q(b)=DF^q(F^n(b))
$$
for any periodic point $b$ and $n\in\N=\{0,1,2,\dots\}$.
Under our assumptions, necessarily
$$
a_1<b_1<a_2<\dots <b_{N-1}<a_N<b_N<F(a_1)
$$
are just the all periodic points in $[a_1,F(a_1)]$.
Writing
\begin{equation}
\int_{a_1}^{F(a_1)}\frac{(f)^2}{DF^{nq}}\, dx  =
\sum_{i=1}^N\int_{J_i}\frac{(f)^2}{DF^{nq}}\, dx
\quad ,
\label{fn}
\end{equation}
all the terms can be treated as in
the proof of Th. 3.25 of \cite{DDG} applied to the function
$F^q$. We know from \cite{DDG} that the sequence of functions
$$
l_n^{(i)}=\prod_{j=0}^{n-1}\frac{DF^q(a_i)}{DF^q\circ F^{jq}}
$$
is uniformly bounded and pointwise convergent in $J_i$
to the strictly positive  function $l^{(i)}$ as $n\to\infty$.
So the limit of
each term in (\ref{fn}) multiplied by $[DF^q(a_i)]^n$ can be
calculated taking the limit under the integral
and the formulas (\ref{nppasympt}-\ref{restasympt}) are obtained.
As $\|f\|_{L^2((a_1,F(a_1)))}>0$ an index $i_0$ surely exists and the
leading term is nontrivial.
\\
\hspace*{13cm} $\fbox{}$

\noindent
{\bf Remark.} The validity of Lemma \ref{asymptotpq} was
mentioned in \cite{DDG} but only the case of integer rotation
number ($\rho(F)=pT$ where $p\in\N^*$) and two periodic points in
$[-a(0),a(0))$ was explicitly written for simplicity. However, the
assumption of only two periodic points then leads to $p=1$
as can be seen using the invariance of the set of periodic points under
$F$ and under the translation by period $T$. We have overlooked this
constraint in \cite{DDG}.

Let us now repeat some assumptions and
formulate some new ones for the purpose of the main theorem.\\

\noindent
{\bf Assumptions.} Let $a\in C^2(\R)$ be a strictly positive
periodic function with a period $T>0$, satisfying $|a'|<1$. Let
the function $F$ defined by relations (\ref{hkFdef}) has the
rotation number
\begin{equation}
\rho(F)=\frac{p}{q}
\end{equation}
where $p,q\in \N^*=\{1,2,\dots\}$ are relatively prime. Let the
function $F$ has a finite number of periodic points in the
interval $I_0=[-a(0),a(0))$, of them $a_1,\dots,a_N$ being
attracting ($N\in \N^*$) and other periodic points being repelling.
Let us denote as $b_0,\dots,b_N$ the repelling periodic points
such that
\begin{equation}
\label{aborder}
b_0<a_1<b_1<\dots<b_{N-1}<a_N<b_N=F(b_0)
\end{equation}
and there are no other periodic points in $(b_0,b_N)$. Notice
that these
repelling periodic points necessarily exist, (\ref{aborder})
is the only possible ordering of periodic points, and just one
of the two periodic points $b_0$, $b_N$ lies in $I_0$.
As the initial data determine the function $G$ directly in the interval
$I_0$ by Eq. (\ref{GI}) while the whole interval
$(b_0,b_1)$ resp. $(b_{N-1},b_N)$ need not be a part of $I_0$ we have to map
the outreaching part into $I_0$ by function $F$ resp. $F^{-1}$ if necessary.
Let us therefore denote
\begin{eqnarray}
\nonumber
J_1&=&(\max\{b_0,-a(0)\},b_1)\cup(\min\{b_N,a(0)\},a(0)) \quad ,\\
J_i&=&(b_{i-1},b_i) \quad {\rm for} \quad i=2,\dots,N-1 \quad ,\\
\nonumber
J_N&=&(b_{N-1},\min\{b_N,a(0)\})\cup(-a(0),\max\{b_0,-a(0)\})
\end{eqnarray}
where notation $(x,y)=\emptyset$ for $x\geq y$ is used. These
formulas can be also written in a more compact and for the
further use clearer way as
$$
J_i=\buildrel{\;\circ}\over I_0\bigcap\bigcup_{j=-1}^1F^j((b_{i-1},b_i))
\quad {\rm for} \quad i=1,\dots,N\;.
$$
Let the all attracting periodic points be such that
\begin{equation}
DF^q(a_i)<1 \quad {\rm for} \quad i=1,\dots,N
\end{equation}
and let index $i_0$ be such that
\begin{equation}
DF^q(a_{i_0})=\min\{DF^q(a_i)|\, i=1,\dots,N\} \quad .
\end{equation}
Let us denote
\begin{equation}
J=\cup\{J_i|\, DF^q(a_i)=DF^q(a_{i_0})\} \quad .
\end{equation}
\\

We are ready to formulate the main statement now.
\begin{theorem}
Let functions $a$, $\varphi_0$ and $\varphi_1$ be as
in Theorem \ref{KGEexist} and satisfy the Assumptions given above.
Let the function $\varphi_0'(|x|) + \varphi_1(|x|)\, {\rm sgn}(x)$ has a
strictly positive norm in $L^2(J)$.
Then there exist strictly positive finite constants $m_0$, $A$ , $B$
(dependent on $\varphi_0$ and $\varphi_1$) such that
for masses $0\leq m\leq m_0$ and time $t\geq 0$ the energy $E_m(t)$ satisfies
the inequality
\begin{equation}
\label{Eexp}
A e^{\gamma t} \leq E_m(t) \leq B e^{\gamma t} \quad
\end{equation}
where
\begin{equation}
\gamma=-\frac{1}{pT}\ln\left(DF^q(a_{i_0})\right)\;>0 \quad .
\end{equation}
\end{theorem}
{\bf Proof.}
Preliminarily, let us show that the terms defining the energy do
not change more than by a constant factor if time undergoes a
constant translation. This will enable us to use a special
sequence of times only. Let us consider two times $t_1$ and $t_2$
satisfying the inequalities
\begin{equation}
\label{tstep1}
h(t_1)\leq h(t_2)\leq k(t_1) \quad ,
\end{equation}
i.e. $t_1\leq t_2\leq h^{-1}\circ k(t_1)$. The last inequality is
clearly satisfied if
\begin{equation}
\label{tstep2}
t_1\leq t_2\leq t_1 + 2 a_{min}
\end{equation}
as the second inequality (\ref{tstep1}) reads $t_2\leq t_1
+ a(t_1) + a(t_2)$. Similar calculations as in the proof of
Lemma 2.17 of \cite{DDG} lead to the estimate for the energy of
massless field
\begin{equation}
\label{E0step}
\frac{1}{F'_{max}}E_0(t_1) \leq E_0(t_2) \leq
\frac{1}{F'_{min}}E_0(t_1) \quad .
\end{equation}
Analogously for
\begin{equation}
S_j(t):=\int_{h(t)}^{k(t)}\left(DF^{-j}(y)\right)^2\,dy
\end{equation}
where $j\in\N$ we can write
\begin{eqnarray*}
S_j(t_1)&=&\left(\int_{h(t_1)}^{h(t_2)}+\int_{h(t_2)}^{k(t_1)}\right)
DF^{-j}(y)^2\,dy \\&=&
\int_{h(t_2)}^{k(t_1)}DF^{-j}(y)^2\,dy +
\int_{k(t_1)}^{k(t_2)}\left(DF^{-j}\circ F^{-1}(y)\right)^2 DF^{-1}(y)\,dy
\\&=&
\int_{h(t_2)}^{k(t_1)}DF^{-j}(y)^2\,dy +
\int_{k(t_1)}^{k(t_2)}DF^{-j}(y)^2
\frac{\left(DF^{-1}\circ F^{-j}(y)\right)^2}{DF^{-1}(y)}\,dy \quad .
\end{eqnarray*}
Estimating the fraction in the last integrand in terms of
$F'_{min}<1<F'_{max}$ and the factor $1$ in the first integral by
the same value we obtain
\begin{equation}
\label{Sstep}
\frac{F'^2_{min}}{F'_{max}}S_j(t_1) \leq S_j(t_2) \leq
\frac{F'^2_{max}}{F'_{min}}S_j(t_1) \quad .
\end{equation}
Let us denote
\begin{equation}
t_0=h^{-1}(a_{i_0}) \quad .
\end{equation}
Combining inequalities (\ref{tstep2}), (\ref{E0step}) and (\ref{Sstep})
we see that for
\begin{equation}
\label{tliken}
npT+t_0\leq t \leq (n+1)pT+t_0
\end{equation}
with any $n\in\N$ the estimates
\begin{eqnarray}
\label{E0bound}
L_1 E_0(npT+t_0)\;\leq&E_0(t)&\leq\; L_2 E_0(npT+t_0) \quad , \\
\label{Sbound}
M_1 S_j(npT+t_0)\;\leq&S_j(t)&\leq\; M_2 S_j(npT+t_0)
\end{eqnarray}
hold where
\begin{eqnarray}
\nonumber
L_1&=&F'^{-n_0}_{max}\, ,\; L_2=F'^{-n_0}_{min}\, ,\;
M_1=\left(\frac{F'^2_{min}}{F'_{max}}\right)^{n_0} \, ,\;
M_2=\left(\frac{F'^2_{max}}{F'_{min}}\right)^{n_0} \, ,
\\
n_0&=&\left[\frac{pT}{2a_{min}}\right] + 1 \,
\end{eqnarray}
the square brackets denoting the entire part here.

Let us denote as
\begin{equation}
\tilde{E}(t)=\sum_{\omega=\xi,\eta}
\left\|\frac{\partial\vp}{\partial\omega}\right\|^2_{L^2((0,a(t)),dx)}
\end{equation}
and let us write Eq. (\ref{deriter}) as
\begin{equation}
\frac{\partial\vp}{\partial\omega} =
\frac{\partial\vp^{(0)}}{\partial\omega} + \kappa_\omega \quad .
\end{equation}
Now
\begin{equation}
\label{E0kappa}
\sqrt{E_0(t)}-\sqrt{\sum_{\omega=\xi,\eta}\|\kappa_\omega\|^2} \leq
\sqrt{\tilde{E}(t)} \leq
\sqrt{E_0(t)}+\sqrt{\sum_{\omega=\xi,\eta}\|\kappa_\omega\|^2}
\end{equation}
by the triangle inequality. By estimates (\ref{Aestim}),
\begin{eqnarray}
\nonumber
\sqrt{\sum_{\omega=\xi,\eta}\|\kappa_\omega\|^2} &\leq&
\frac{m^2}{4}c_2(m)\sum_{j=0}^{K(t)}e^{(K(t)-j)a_{max}^2m^2}
\|DF^{-j}\|_{L^2(h(t),k(t))} \\&\leq&
\frac{m^2}{4}c_2(m)\sqrt{M_2}\sum_{j=0}^{K(t)}e^{(K(t)-j)a_{max}^2m^2}
\sqrt{S_j(npT+t_0)} \quad
\label{kappanorm}
\end{eqnarray}
for $t$ satisfying (\ref{tliken}). For $j=iq+r$ with $i\in\N$ and
$r=0,\dots,q-1$ let us calculate
\begin{eqnarray}
\nonumber
S_j(npT+t_0)=\int_{npT+a_{i_0}}^{npT+F(a_{i_0})}DF^{-j}(y)^2\,dy
= \int_{a_{i_0}}^{F(a_{i_0})}DF^{-j}(y)^2\,dy
\\ \nonumber
= \int_{F^{-j}(a_{i_0})}^{F^{1-j}(a_{i_0})}DF^{-j}\circ F^j(y)\,dy
=  \int_{F^{-r}(a_{i_0})}^{F^{1-r}(a_{i_0})}\frac{dy}{DF^j(y)}
\\ =
\int_{a_{i_0}}^{F(a_{i_0})}\frac{DF^{-r}(y)^2}{DF^{j-r}(y)}\,dy
\leq F'^{-2r}_{min} \int_{a_{i_0}}^{F(a_{i_0})}\frac{dy}{DF^{iq}(y)}
\quad .
\end{eqnarray}
By Lemma \ref{asymptotpq} there exists a finite constant $c_3>0$ such
that
\begin{equation}
0 < \int_{a_{i_0}}^{F(a_{i_0})}\frac{dy}{DF^{iq}(y)} < c_3
[DF^q(a_{i_0})]^{-i} \leq c_3 [DF^q(a_{i_0})]^{-\frac{j}{q}} \quad .
\end{equation}
Estimate (\ref{kappanorm}) now gives
\begin{eqnarray}
\nonumber
\sqrt{\sum_{\omega=\xi,\eta}\|\kappa_\omega\|^2} \leq
\phantom{xxxxxxxxxxxxxxxxxxxxxxxxxxxxxxxxxxxxxx}
\\
\frac{m^2}{4}c_2(m)\sqrt{c_3 M_2}F'^{-(q-1)}_{min}
e^{K(t)a_{max}^2m^2}
\frac{\left(DF^q(a_{i_0})^{-\frac{1}{2q}}e^{-a_{max}^2m^2}\right)^{K(t)+1} - 1}
{DF^q(a_{i_0})^{-\frac{1}{2q}}e^{-a_{max}^2m^2} - 1} .
\end{eqnarray}
Let us now choose $m_1>0$ such that
\begin{equation}
m_1^2 < -\frac{1}{2qa_{max}^2}\ln\left(DF^q(a_{i_0})\right)
\end{equation}
and consider only mass values
\begin{equation}
0\leq m \leq m_1 \quad .
\end{equation}
Now we can estimate
\begin{equation}
\sqrt{\sum_{\omega=\xi,\eta}\|\kappa_\omega\|^2} \leq
c_4 m^2 \left(DF^q(a_{i_0})\right)^{-\frac{K(t)+1}{2q}}
\end{equation}
where we denoted
\begin{equation}
c_4=
\frac{\frac{1}{4}c_2(m_1)\sqrt{c_3 M_2}F'^{-(q-1)}_{min}}
{DF^q(a_{i_0})^{-\frac{1}{2q}}e^{-a_{max}^2m_1^2} - 1}
e^{-a_{max}^2m_1^2}
\quad .
\end{equation}

The energy of the massless field is given by function $G_0$
defined by Eqs. (\ref{GI}-\ref{prolong}) with $f=0$. For the
considered sequence of times it reads
\begin{eqnarray}
\nonumber
E_0(npT+t_0) &=& \int_{npT+a_{i_0}}^{npT+F(a_{i_0})}G_0'(y)^2\,dy
\\ &=&
\int_{F^{nq}(a_{i_0})}^{F^{nq+1}(a_{i_0})}G_0'(y)^2\,dy =
\int_{a_{i_0}}^{F(a_{i_0})}\frac{G_0'(y)^2}{DF^{nq}(y)}\,dy
\; .
\end{eqnarray}
Now Lemma
\ref{asymptotpq} and inequality (\ref{E0bound}) show the existence
of constants $0<D_1<D_2<\infty$ such that
\begin{equation}
\label{E0nbound}
D_1 \left(DF^q(a_{i_0})\right)^{-n} \leq E_0(t)
\leq D_2 \left(DF^q(a_{i_0})\right)^{-n}
\end{equation}
for $t$ satisfying (\ref{tliken}) as it is clear from
the prolongation formula for $G_0$, (\ref{prolong}) with $f=0$, that
the assumed nontriviality of
$\varphi_0(|x|)+\varphi_1'(|x|)\,{\rm sgn}(x)$ in $J$
leads to the nontriviality of $G_0'$ in a suitable neighborhood
of a suitable attracting periodic point of $F$ in
$[a_{i_0},F(a_{i_0}))$.

Let us relate the numbers $n$ in (\ref{tliken}) and $K(t)$ defined
by (\ref{Ktdef}) and (\ref{ndef}). As (\ref{tliken}) gives $$
t+a_{max}\geq k(t)=F(h(t))\geq F(npT+a_{i_0})\geq F^{nq+1}(-a(0))
$$ we have \begin{equation} K(t)\geq nq+1 \quad . \end{equation}
Similarly (\ref{tliken}) implies \begin{eqnarray*}
t+a_{max}=h(t)+a(t)+a_{max}\leq h(t)+2a_{max} \\ \leq a_{i_0} +
(n+1)pT + 2a_{max} <a_{i_0} +
\left(n+\left[\frac{2a_{max}}{pT}\right]+2\right)pT \\ <
F^{\left(n+\left[\frac{2a_{max}}{pT}\right]+2\right)q}(a(0))
\end{eqnarray*} where square brackets denotes the entire part and
\begin{equation} K(t)\leq
\left(n+\left[\frac{2a_{max}}{pT}\right]+2\right)q \quad .
\end{equation}

Estimate (\ref{E0nbound}) can be written as
\begin{equation}
\label{E0Kbound}
D'_1 [DF^q(a_{i_0})]^{-\frac{K(t)}{q}} \leq E_0(t)
\leq D'_2 [DF^q(a_{i_0})]^{-\frac{K(t)}{q}}
\end{equation}
with
\begin{equation}
D'_1=D_1\left[DF^q(a_{i_0})\right]^{\left[\frac{2a_{max}}{pT}\right]+2}
\quad , \quad D'_2=D_2\left[DF^q(a_{i_0})\right]^{\frac{1}{q}}
\quad .
\end{equation}
Let us choose $m_2$ such that
\begin{equation}
0< m_2 <c_4^{-\frac{1}{2}} D'^{\frac{1}{4}}_1
\left[DF^q(a_{i_0})\right]^\frac{1}{4q}
\end{equation}
and denote
\begin{eqnarray}
\nonumber
C_1&=&\left(\sqrt{D'_1} -
c_4m_2^2\left[DF^q(a_{i_0})\right]^{-\frac{1}{2q}}\right)^2
\quad , \\
C_2&=&\left(\sqrt{D'_2} +
c_4m_2^2\left[DF^q(a_{i_0})\right]^{-\frac{1}{2q}}\right)^2
\quad .
\end{eqnarray}
Then by (\ref{E0kappa})
\begin{equation}
\label{EtildbK}
C_1 \left[DF^q(a_{i_0})\right]^{-\frac{K(t)}{q}} \leq
\tilde{E}(t) \leq C_2 \left[DF^q(a_{i_0})\right]^{-\frac{K(t)}{q}}
\end{equation}
for any $t\geq 0$ and
$0\leq m \leq \min(m_1,m_2)$.

Let us relate $K(t)$ to $t$. For $t\geq 0$,
$$
t+a_{max}\geq a(0) = F(-a(0))
$$
and therefore $K(t)\geq 1$. For any $x\in\R$ and $n\in\N^*$ the
relation
\begin{equation}
\nonumber
-\frac{T}{n}<\frac{F^n(x)-x}{n}-\rho(F)<\frac{T}{n}
\end{equation}
holds by Proposition II.2.3 of Ref. \cite{Herman} (the same relation was
used in the proof of Lemma 2.17 in \cite{DDG}). Putting here
$n=K(t)$, $x=F^{-n}(t+a_{max})$,
$\rho(F)=\frac{p}{q}T$ and taking into account
definitions (\ref{Ktdef}), (\ref{ndef}) we obtain
\begin{equation}
\frac{t}{pT} + \frac{a_{max}-a(0)-T}{pT} < \frac{K(t)}{q} <
\frac{t}{pT} + \frac{a_{max}+a(0)+T}{pT}
\quad .
\end{equation}
Now (\ref{EtildbK}) gives
\begin{equation}
C'_1 e^{\gamma t} \leq \tilde{E}(t) \leq C'_2 e^{\gamma t}
\end{equation}
with
\begin{equation}
C'_1=C_1\left[DF^q(a_{i_0})\right]^{\frac{T+a(0)-a_{max}}{pT}}
\; , \;
C'_2=C_2\left[DF^q(a_{i_0})\right]^{-\frac{T+a(0)+a_{max}}{pT}}
\; .
\end{equation}

Let us estimate the contribution of the mass term to the energy
(\ref{energydef}). By the estimate (\ref{phibound}),
\begin{equation}
\int_0^{a(t)} \vp(\xi,\eta)^2\,dx \leq
c^2 a_{max} e^{a_{max}m^2(t+a_{max})}
\; .
\end{equation}
If we now denote
\begin{eqnarray}
\label{m0def}
m_0=\min\left(m_1,m_2,\sqrt{\frac{\gamma}{a_{max}}}\right) \quad ,
\\
A=C'_1 \quad , \quad
B=C'_2 + \frac{1}{2}c^2 a_{max} m_0^2 e^{a_{max}^2m_0^2}
\end{eqnarray}
the estimate (\ref{Eexp}) follows for every $0\leq m\leq m_0$ and
$t\geq 0$.
\\
\hspace*{13cm} $\fbox{} $
\\
{\bf Remark.} If the condition $m\leq \sqrt{\frac{\gamma}{a_{max}}}$
is relaxed from (\ref{m0def}) we have still exponential lower and
upper bounds for
the energy time-development but the upper exponent may be higher
than that for the massless case. However, we cannot claim that
such bound would be saturated as we do not know whether the
estimate (\ref{phibound}) can be improved substantially. In
particular, we do not know whether the field $\varphi$ is bounded
as in the massless case since we were able to prove the
exponential estimate only.
\\ \\
{\bf Acknowledgments.}
The authors are indebted to Prof. J. Cooper and Dr. N. Gonzalez
for discussions. The stages of J.D. at CPT CNRS and PhyMat UTV
and P.D. at NPI ASCR are thankfully acknowledged. The work is
partly supported by ASCR grant No. IAA1048101.

\end{document}